\def\kms{\hbox{km s$^{-1}$}}

\def\hii{H{\sc ii}}
\def\msun{M$_\odot$}
\def\mjyb{\hbox{mJy beam$^{-1}$}}
\def\msunyr{\hbox{M$_\odot$ yr$^{-1}$}}

\def\cm{cm$^{-3}$}

\def\gra{$^{\circ}$}

\def\Ha{H${\alpha}$}
\def\deg{$^{\circ}$}
\def\radec{RA, Dec.(J2000)}
\def\radecb{RA, Dec.(J2000)}
\def\micr{$\mu$m}

\documentclass[structabstract]{aa}  

\usepackage{graphicx}
\usepackage{txfonts}
\usepackage{natbib}

\begin{document}
\title{Molecular gas towards G18.8+1.8 }

\author{J. Vasquez\inst{1,2}\thanks{Member of 
Carrera del Investigador, CONICET, Argentina}
\and 
M. Rubio\inst{3} 
\and 
C.E. Cappa\inst{1,2 \star}
\and
N. U. Duronea\inst{2}
}
\institute{Facultad de Ciencias Astron\'omicas y Geof\'{\i}sicas,  Universidad Nacional de La Plata, Paseo del Bosque s/n, 1900 La Plata, Argentina \email{pete@iar.unlp.edu.ar} \and Instituto Argentino de Radioastronom\'{\i}a, C.C. 5, 1894 Villa Elisa, Argentina \and Departamento de Astronom\'{\i}a, Universidad de Chile, Casilla 36-D, Santiago, Chile}

\date{Received September 15, 1996; accepted March 16, 1997}

 
\abstract
    {}
    {This work aims at investigating the characteristics of the molecular gas associated with the nebula G18.8+1.8, linked to the Wolf-Rayet star HD\, 168206 (WR 113), ant its relation to other components of its local interstellar medium.}
     {We carried out molecular observations of the  $^{12}$CO($J=1-0$) and ($J=2-1$) lines with angular resolution of 44\arcsec\ and 22\arcsec using the SEST telescope. Complementary NANTEN data of the $^{12}$CO(1-0) line were also used. The dust emission was analyzed using Spitzer-IRAC images at 8.0 $\mu$m, and WISE data at 3.4 $\mu$m, 4.6 $\mu$m, and 12.0 $\mu$m.}
     {The SEST data allowed us to identify a molecular component (Cloud 3) having velocities in the  interval from $\sim$ +30 to +36 \kms\    which is most probably linked to the nebula. Morphological and kinematical properties suggest that Cloud 3 constitute a wind-blown molecular half-shell, which expands around WR 113. The ratio $R_{2-1/1-0}$ and excitation temperatures  indicate that the molecular gas is being irradiated by strong  UV radiation. The location of the inner optical ring in the outer  edge of Cloud 3 suggests that the stars SerOB2-1, -2, -3, -63, and -64 are  responsables for the ionization of Cloud 3 and the inner ring nebula. A comparison between the spatial distribution of the molecular gas and the PAH emission at 8 $\mu$m indicates the existence of  a PDR between the ionized and the molecular gas.

A search for candidate  young stellar objects (YSOs) in the region around G18.8+1.8 based on available 2MASS, MSX, IRAS, and {\it Spitzer}-IRAC  catalogs resulted in the detection of about  sixty sources, some of them projected onto Cloud 3. Two small spots of clustered candidates YSOs are projected near the outer border of Cloud 3, although a triggered stellar formation scenario is doubtful.  }
{}


\titlerunning{The interstellar medium in the environs of WR\,113}

\maketitle
%

\section{Introduction}\label{section1}

Most massive star formation in our Galaxy occurs in Giant Molecular clouds (GMC) where, in general, the star formation is developed in groups like stellar clusters or OB associations. These  groups can contain hundreds of OB stars, including massive young stellar objets (MYSOs) and evolved massive stars. Since OB associations directly influence the evolution of their environs, it is important to study how this interaction occurs in different interstellar environments. Massive stars interact with the interstellar medium  (ISM) through their intense ultraviolet (UV) radiation field ($h\nu>$ 13.6 eV) and via strong stellar winds. UV photons ionize the surrounding gas creating \hii\ regions and dissociate the molecular gas originating photodissociation regions (PDRs), while the second mechanism alters the structure of \hii\ regions sweeping up the circumstellar material and creating interstellar bubbles (IB) \citep{ca05}.\

Ser\,OB2 is a rich OB association  located at $\radec \simeq$ (18$^h$18$^m$00$^s$, --11\gra36\arcmin36\arcsec), at a distance of 1.9 $\pm$ 0.3 kpc \citep{f00}, about 70 pc above the Galactic plane. \citet{f00} performed a very interesting spectrophotometric study of this association and analyzed its connection to a narrow column of hot gas emerging from the \hii\ region Sh2-54 perpendicularly to the Galactic plane, known as the thermal ``chimney'' \citep{mu87}. From the original set of stars ($\sim$500) taken for the analysis, only 107 were considered to be probable members of Ser\,OB2 (see Table 5 from \citealt{f00}). Ser OB2 contains a number of massive stars including the eclipsing binary system CV\,Ser \citep{mn81}, the Of-type binary HD 166734 \citep{co00}, the O-type multiple star HD 167971 (MY Ser ; \citealt{le87}), and the O5.5Vf star HD 168112.

   Our aim in this paper is to investigate the existence of molecular gas in the environs of CV\,Ser and other posible members of Ser\,OB2  based on high angular resolution SEST observations performed in the $^{12}$CO(1-0) and $^{12}$CO(2-1) lines, complemented with    medium angular resolution CO (1-0) data from the NANTEN telescope and  infrared images in the mid infrared (MIR). 

\subsection{Background of CV\,Ser and its close environs}\label{sc}

\begin{figure*}
\centering
\includegraphics[angle=0,width=183mm]{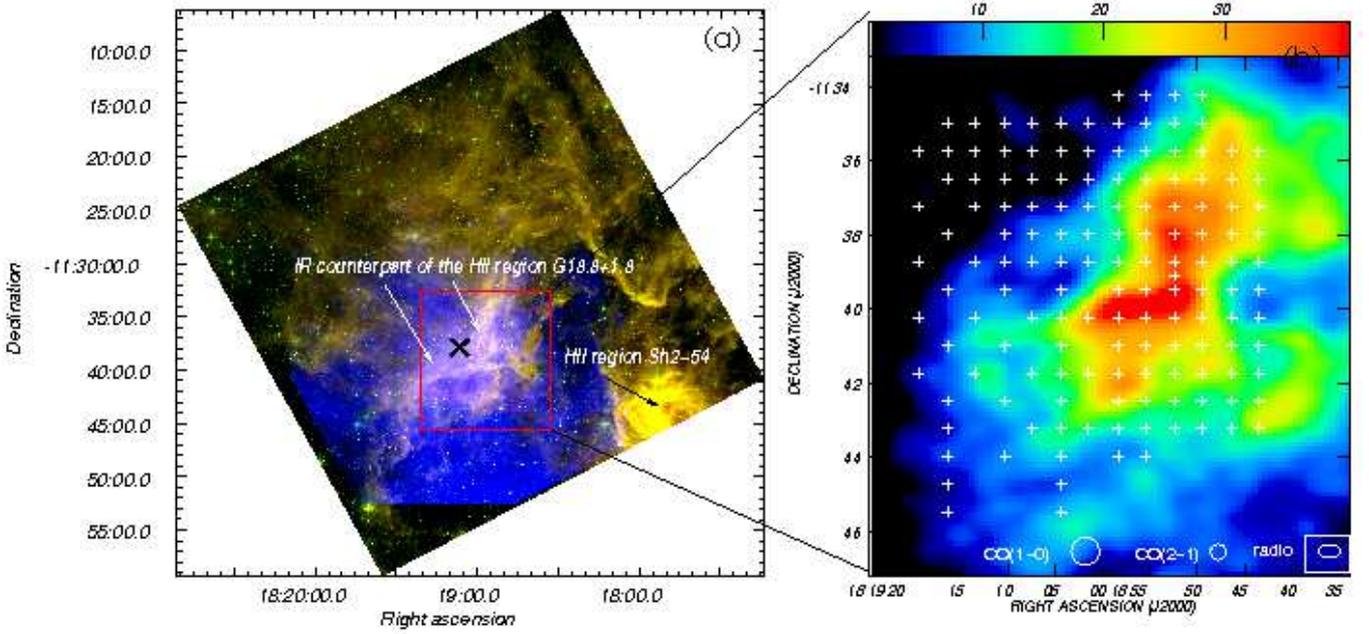}
\caption{{\it Left panel:} Superposition of the IRAC images at 8.0 $\mu$m (red) and 5.6$\mu$m (green) with the \Ha\ emission (blue), of a section of the \hii\ region Sh2-54 including the environs of WR\,113. The cross marks the position of the WR star. {\it Right panel:} VLA image at 1465 MHz showing the ionized ring. The crosses mark the positions of the 143 points observed in $^{12}$CO lines. The colour scale goes from 1 to 40 \mjyb. The synthesized beam of the radio continuum observations and the beams of the molecular data are indicated in the bottom part of the image.}
\label{opt1}
\end{figure*}

 The brightest component of   CV\,Ser is the Wolf-Rayet star WR\,113 ($\equiv$ HD\,168206, WC8d+O8-9IV, \citealp{vdh01}).  Table~\ref{tabla0} summarizes the main parameters of WR\,113: coordinates, spectral classification, visual absorption $A_v$, spectrophotometric distance, mass loss rate $\dot{M}$, and terminal velocity v$_{\infty}$.

     \citet{gr84} discovered a double optical structure of 4\arcmin\ and 9\arcmin\ in radius around the WR star and suggested that the outer ring, which is poorly defined, was formed by the WR star or its massive progenitor and is presently being photoionized by the O5.5Vf star HD\,168112 (at \radec\ = \hbox{(18$^h$18$^m$40.8$^s$,--12\gra 06\arcmin23.4\arcsec)}, also belonging to Ser OB2 and located to the southwest of the ring nebula. \citet{er95} observed the inner ring in [N{\sc ii}] $\lambda$$\lambda$6548,6584, [S{\sc ii}] $\lambda$$\lambda$6717,6731, \Ha\ $\lambda$6563, and H$\beta$\ $\lambda$4861. They determined that the velocity of the ionized gas is in the range [$\sim$+27,$\sim$+39] \kms, which corresponds to a kinematical distance of 2.0 $\pm$ 0.2 kpc, according to circular galactic rotation models (e.g. \citealp{bb93}). A similar velocity of +27 \kms\ was found from radio recombination lines by \citet{lo89}. \citet{er95} concluded that photoionization is the main source of excitation of the inner ring, being the WR star the main responsible. Following the classification by \citet{chi91} for WR ring nebulae, they classified the inner nebula as $R_s$ type, because it does not present  evidence of expansion and its morphology is shell-structured.

The left panel of Fig.~\ref{opt1} shows a superposition of the IRAC images at 8.0 $\mu$m (in red) and 5.6 $\mu$m (green), with the \Ha\ emission (blue) of the ring nebula linked to CV\,Ser. The strong emission at  \radec\ = \hbox{(18$^h$18$^m$00$^s$,--11\gra 44\arcmin)} corresponds to the \hii\ region Sh2-54. The bright filament  to the west and south of the  binary system, which is indicated by the cross symbol, corresponds to the inner structure of 4\arcmin\ in radius. This structure displays strong emission both in the \Ha\  line and in the mid infrared.

The right panel of  Fig.~\ref{opt1} displays the radio continuum emission distribution at 1465 MHz, as observed by \citet{c02} using the Very Large Array (VLA) in DnC configuration with a synthesized beam of 30\arcsec. The radio continuum source was first identified as G18.8+1.8 by \citet{gd70} at 2.7 GHz. The high angular resolution VLA data allowed to detect radio counterparts of both the inner and outer rings. Adopting a distance of 2.0 $\pm$ 0.6 kpc, \cite{c02} obtained ionized masses and electron densities of 20 $\pm$ 10 \msun\ and  180 - 500 \cm, respectively, for the inner shell, and 90 \msun\ and 40 \cm, respectively, for the outer shell. From the IR-radio continuum relation, they confirmed the thermal nature of the nebula. Adopting a radius $R_s$ = 2.3 pc for the inner ring and an expansion velocity $v_{exp}$ = \hbox{5 - 10} \kms, the derived dynamical age $t_d$ = \hbox{(1.3 - 2.5) $\times$ 10$^5$} yr suggests that the inner shell originated in the action of the stellar wind of the current WR phase.

A search for OB stars within a region 12$'$ in size centered on the WR star  indicated that other excitation sources are present in this region in addition to the binary system. Ser\,OB2-1,-2, and -3 are confirmed members of the OB association, while Ser\,OB2-63, and -64 are potencial members (see Table 5 from  \citealt{f00}).    Their coordinates and (B-V) and (U-B) colors  are listed in Table\ref{tabla1}. \

In our study, in agreement with \citet{f00}, we adopt a distance of 2.0 $\pm$ 0.6 kpc. This value is also in agreement with the cataloged distances of the WR star (see Table \ref{tabla0})
 
\begin{table}
\begin{center}
\caption{Main parameters of WR\,113}
\label{tabla0}
\begin{tabular}{lc}
\hline
\radec\ & (18$^h$19$^m$7.36$^s$ ,--11\gra37\arcmin 59\farcs 20) \\
Spectral classification & WC8d+O8-O IV \\
$A_v$ (mag) & 3.23$\pm$0.1$^a$ \\
$d$ (kpc) &  1.8$^a$, 2.0$^b$, 2.5$^{e,f}$ \\
{\bf $\dot{M}$ (10$^{-5}$ \msunyr\ }) &$<$ 5.6$^g$, 2.0$\pm$0.3$^h$, 2.4$^f$ \\
v$_{\infty}$ (\kms)\ & 1400$^i$, 1890$^j$ \\
\hline 
\end{tabular}
\end{center}
\footnotesize{
Notes: $(a)$ \citet{vdh01}, $(b)$ \citet{cv90}, $(e)$ \citet{er95}, $(f)$ \citet{nl00}, $(g)$ \citet{lei97}, $(h)$ \citet{lam96}, $(i)$ \citet{kh95}, $(j)$ \citet{nug98}.}
\end{table}
\begin{table}
\centering
\caption{Other excitation sources identified in the environs of CV\,Ser.}
\label{tabla1}
\begin{tabular}{lcccc}
\hline
      &RA $J(2000)$  &    DEC.$J(2000)$ & $(B-V)$  & $(U-B)$\\
      &              &                  &               &        \\
\hline
Ser\,OB2-1 & 18 18 36.9 &--11 40 57.8 & 1.32 & 0.01 \\
Ser\,OB2-2 & 18 18 39.9 &--11 43 07.3 & 1.27 & 0.06 \\
Ser\,OB2-3 & 18 18 42.4 &--11 43 54.8 & 1.25 & 0.11 \\
Ser\,OB2-63 & 18 18 53.4&--11 44 10.1 & 0.45 & 0.30 \\
Ser\,OB2-64 & 18 19 05.0&--11 42 58.3 & 0.81 & 0.22 \\
\hline
\end{tabular}
\footnotesize{\\ Notes: from \citet{f00}.}
\end{table}


\section{SEST observations and complementary data}\label{data}

High angular resolution \hbox{$^{12}$CO(J = 1$\rightarrow$0)} and \hbox{$^{12}$CO(J = 2$\rightarrow$1)} data at 115 and 230 GHz, respectively, were obtained with the 15m Swedish-European Submillimitre Telescope (SEST), at La Silla, Chile during two observing runs in February 2002 and March 2003. 
The half-power beam-width of the telescope was 44\arcsec\  and 22\arcsec\
at 115 and 230 GHz, respectively. A high resolution acousto-optical 
spectrometer was used. It consisted of 1000 chanels, with a total bandwidth 
of 100 MHz and a resolution of 40 KHz, giving velocity resolutions of 
0.105 \kms\ at 115 GHz and 0.052 \kms\ at 230 GHz. The system temperatures 
were $\approx$ 400 K at 230 GHz and $\approx$ 320 K at 115 GHz.
Pointing was checked once during each observing run on the SiO (v=1, 
$J$=2$\rightarrow$1) source VX Sgr. Pointing errors were 3\arcsec. 
The uncertainty in the intensity calibration was 10\%. Details about the 
telescope and receivers can be found in  \citet{bo89}.\

The $^{12}$CO(2-1) and $^{12}$CO(1-0) data were acquired simultaneously in the 
position-switching mode on a grid with a spacing of 45\arcsec. The off-source 
position, at which no CO emission was detected,  was  placed at 
\radec\ = \hbox{(18$^h$19$^m$15$^s$, --11\deg 44\arcmin)}. 
A total of 143 $^{12}$CO spectra were taken towards the inner ring and its 
environs. The observed positions are indicated by crosses on the VLA image 
(Fig. \ref{opt1}, right panel).

The spectra were reduced using the CLASS software (GILDAS working 
group)\footnote{htto://www.iram.fr/IRAMFR/PDB/class/class.html}. The line intensities are expressed as main-beam brightness temperatures 
$T_{mb}$, by dividing the antena temperature $T^{\ast}_A$ by the main-beam 
efficiency $\eta_{mb}$, equal to 0.72 and 0.57 at 115 and 230 GHz, 
respectively \citep{jo98}.\ The angular resolution of the $^{12}$CO(2-1) and $^{12}$CO(1-0) spectra, 
the velocity range, the original velocity resolution, the velocity resolution 
after smoothing, and the typical $rms$ noise temperatures are listed in Table \ref{tabla-sest}.
$^{12}$CO(2-1) and $^{12}$CO(1-0) data cubes were constructed using
AIPS software.

Complementary $^{12}$CO(1-0) data obtained with the 4-m {\rm NANTEN} millimeter-wave telescope of Nagoya University were used to investigate the large scale distribution of the molecular gas  in a large area including SerOB2 and the WR star. The half-power beamwidth was  2\farcm 6.  The 4K cooled SIS mixer receiver provided typical system temperatures of $\approx$ 220K (SSB) at this frecuency. The acoustoptical spectrometer (AOS) provided a velocity coverage  range of 100 \kms\ and a velocity resolution of 1.0 km/s.

The distribution of the IR emission was analyzed using Spitzer images  at 5.6 and 8.0 $\mu$m from the Galactic Legacy Infrared Mid-Plane Survey Extraordinaire 
(Spitzer-GLIMPSE, \citealp{b03}) retrieved from the Spitzer Science Center\footnote{http://scs.spitzer.caltech.edu}.  Images at 3.4, 4.6, and 12.0 $\mu$m with angular resolutions of
 6\farcs 1, 6\farcs 5, and  6\farcs 5 from the Wide-field Infrared Survey Explorer (WISE; \citealt{wr10}) were retrieved from IPAC\footnote{http://www.ipac.caltech.edu}.
\begin{table}
\caption{Observational parameters of the  $^{12}$CO(1-0) and $^{12}$CO(2-1) lines observed with SEST.}
\label{tabla-sest}
\begin{tabular}{l c c} 
\hline
                        &  $^{12}$CO(2-1) &  $^{12}$CO(1-0)      \\
\hline
Angular resolution (\arcsec)     &   22        &  45  \\
Velocity range (\kms)            &  (--7,+47) &  (--40,+70)\\
Original velocity                 &  0.054  &  0.110  \\
resolution (\kms)                 &         &   \\
Velocity resolution    &  0.221  & 0.332 \\
 after smoothing (\kms)  &     &  \\
{\it rms} noise (K)  &   0.17  & 0.25  \\
\hline
\end{tabular}
\end{table}

\section{The characteristics of the molecular gas}\label{morfologia}

\subsection{Spatial distribution}

\begin{figure}
\centering
\includegraphics[angle=270,width=60mm]{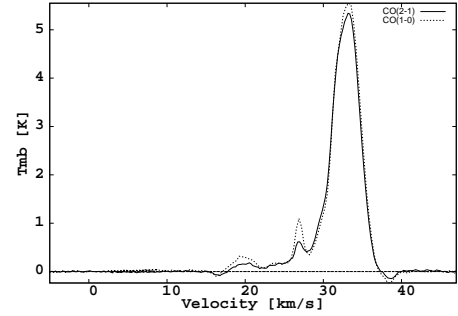}
\caption{$^{12}$CO(2-1) and $^{12}$CO(1-0) spectra averaged within the 
surveyed region. Line intensities are expressed as main-beam brightness temperature $T_{mb}$. Velocities are referred to the LSR.}
\label{vel-cubos}
\end{figure}

\begin{figure*}
\centering
\includegraphics[angle=0,width=125mm]{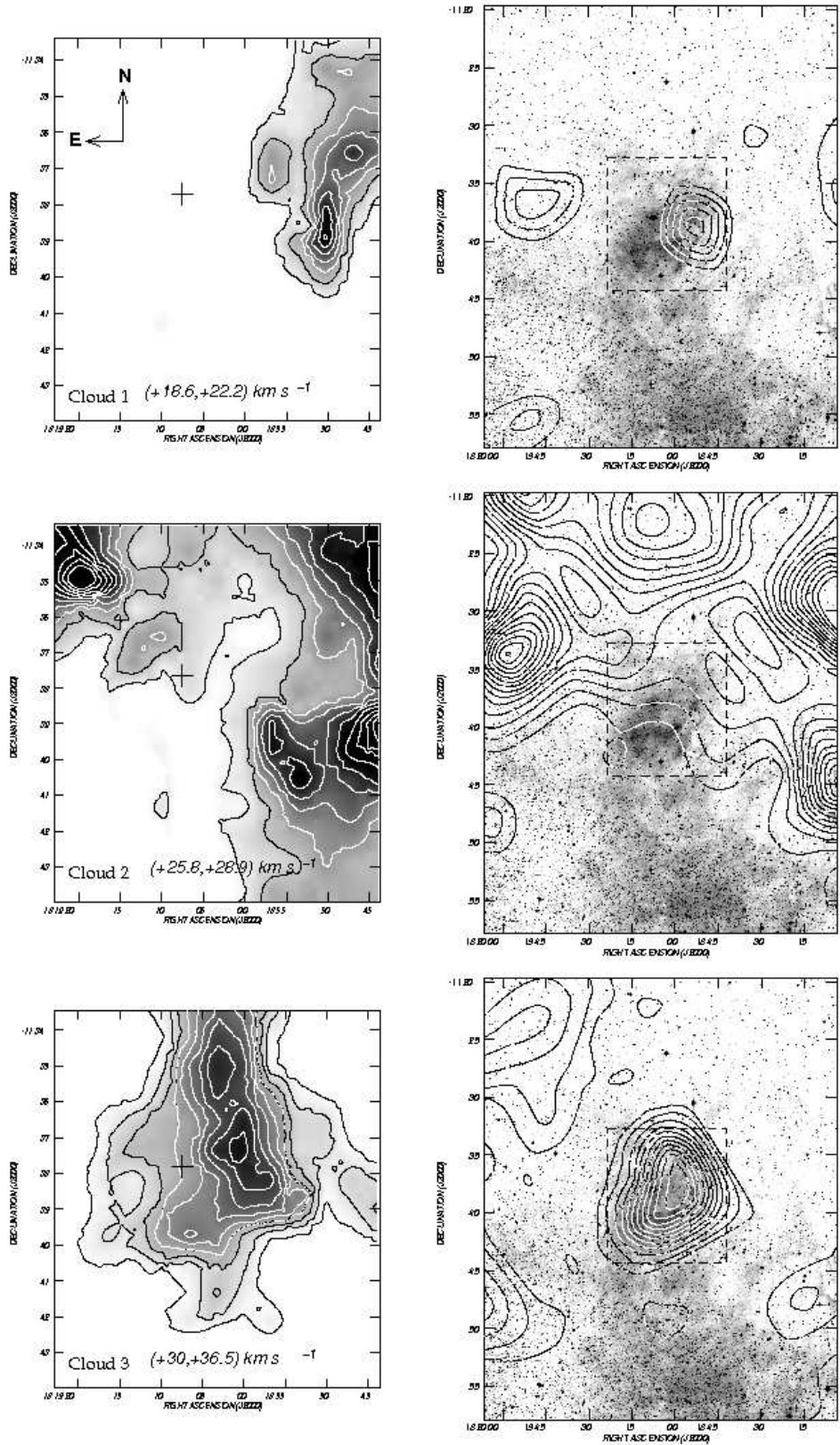}
\caption{{\it Left panels}: $^{12}$CO (1-0) emission distribution maps corresponding to Cloud 1 to 3 obtained with the SEST data. For Cloud 1, the grayscale goes from 0.2 to 2 K, and the contour levels are from 0.3 K ($\sim$ 4.5 $rms$) in steps of 0.3 K. For the case of Cloud 2, the grayscale goes from 0.2 to 0.25 K, and the contour levels are from 0.3 K ($\sim$ 4 $rms$) in steps of 0.3 K. For Cloud 3, the grayscale goes from 1 to 8 K, and the contour levels goes from 1 K ($\sim$ 20 $rms$) in steps of 1 K. The cross indicates the position of the WR star.  {\it Right panels:} $^{12}$CO (1-0) emission distribution maps corresponding to Clouds 1 to 3 obtained with the  NANTEN data (contours), superimposed onto the DSSR image (grayscale). The dashed rectangles at the center of the field indicate the area of the SEST images at the left.     }
\label{nubes1}
\end{figure*}

Aimed to detect the molecular gas possibly linked to the ring nebula, we have inspected the  data cubes of the $^{12}$CO(2-1) and $^{12}$CO(1-0) lines obtained with SEST. The $^{12}$CO(2-1) and $^{12}$CO(1-0) spectra averaged within the surveyed region are displayed in Fig.~\ref{vel-cubos}. Carbon monoxide emission higher than 5 $rms$  is detected between +18 and +37 \kms. Three molecular components are noticed within  the mentioned velocity interval, peaking at  $\sim$ +20  \kms, $\sim$+ 28 \kms, and +34 \kms.    In Fig. \ref{nubes1} (left panels) we show the spatial distribution of the molecular gas in the velocity ranges from +18.6 to +22.2 \kms\ (Cloud 1),  from +25.8 to +28.9 \kms\ (Cloud 2), and from +30 to +36.8 \kms (Cloud 3)  obtained from SEST data. CO intensities are expressed as main-beam brightness temperature averaged within the selected velocity intervals. The cross indicates the position of the WR star.   In order to illustrate the large scale distribution of these molecular clouds,  the molecular emission in the same velocity ranges as obtained with the NANTEN data is shown in the right panels of Fig. \ref{nubes1}, superimposed onto the DSSR image.

Cloud 1  is elongated in the north-south direction with  its brightest emission at \radec\ $\sim$ \hbox{(18$^h$18$^m$50$^s$,--11\deg 39\arcmin)}. A fainter clump is also detected at  \radec\ $\sim$ (18$^h$18$^m$57$^s$,\hbox{--11\deg 37\arcmin)}.   The NANTEN image shows that this cloud is projected onto  the western border of the optical nebula and   has a very small angular extension. Using the analytical fit to the circular galactic rotation model  of \citet{bb93} along \hbox{{\it l} $\approx$ 20\gra} we derived for Cloud 1 near and far  kinematical distances of about 2 and 14 kpc, respectively. Bearing in mind the lack of morphological agreement with the optical nebula and its small angular size, we believe that Cloud 1 may be a background object, unrelated to the star and the ionized regions. A distance as far as  $\sim$ 14 kpc can not be ruled out.

\begin{figure*}
\centering
\includegraphics[angle=0,width=178mm]{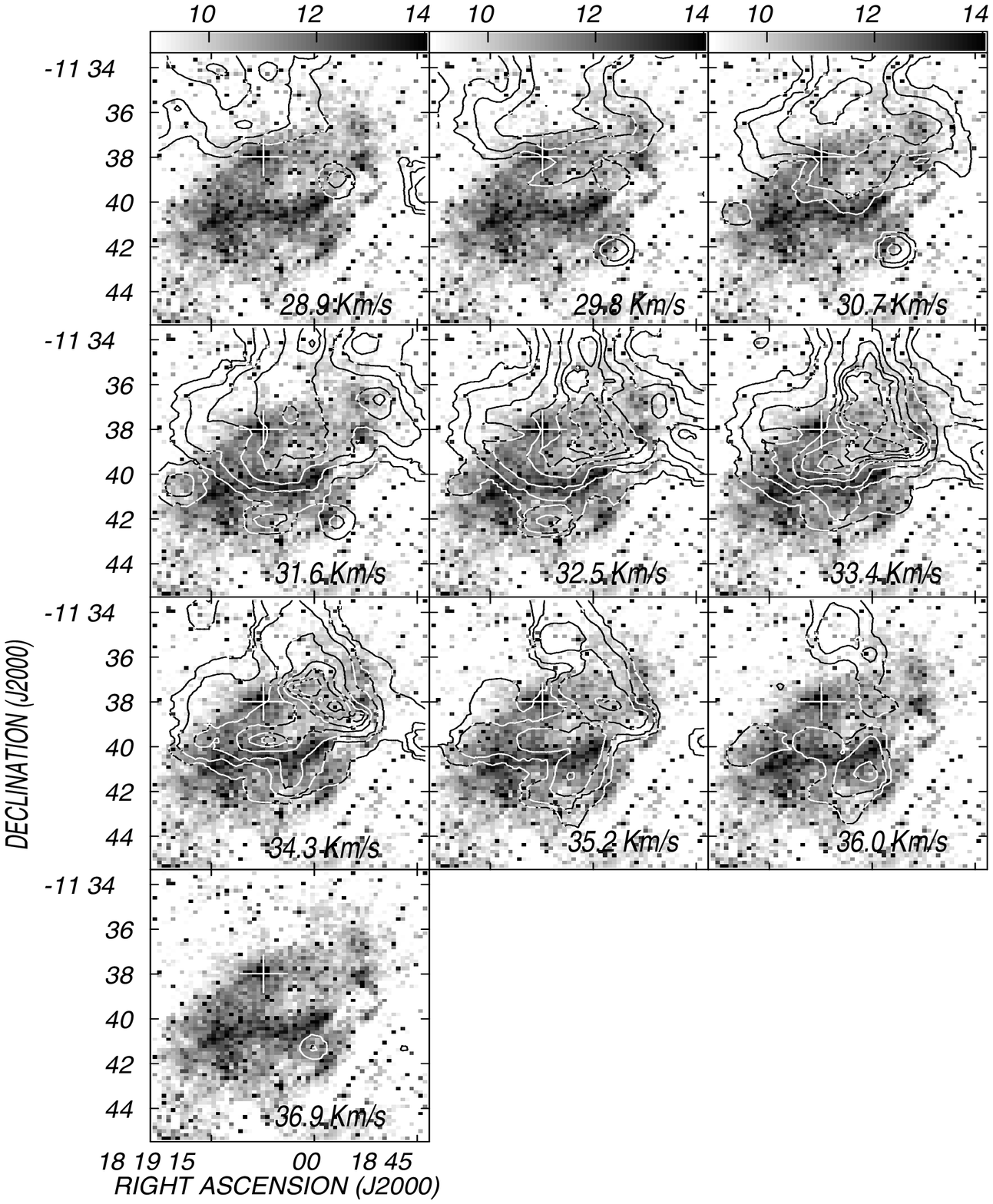}
\caption{Overlay of the mean CO emission (contours) in the velocity range from  28.9 to 36.9 \kms\  and the DSSR emission of the optical nebula (grayscale). The central  velocity is indicated in the lower right corner of each image. Contour levels are 1, 2, 4, 8, 11, 13, and  15 K. The position of WR 113 is indicated with the white cross.  }
\label{fig:serie-1}
\end{figure*}

Cloud 2 displays a  very clumpy morphology. Although it encircles most of the optical nebula, it  shows no clear morphological agreement with the ionized gas. Its maxima are present at  \radec\ $\sim$ (18$^h$18$^m$55$^s$,\hbox{--11\deg 40\arcmin)},  \radec\ $\sim$ \hbox{(18$^h$19$^m$13$^s$,--11\deg 37\arcmin 30\arcsec)}, and \radec\ $\sim$ \hbox{(18$^h$19$^m$20$^s$,--11\deg 35\arcmin)}. Cloud 2 matches with regions lacking optical emission towards the northern, western and southwestern sectors of the nebula. An inspection of the middle right panel of Fig. \ref{nubes1}   clearly indicates that Cloud 2 extends well beyond the area surveyed with SEST and NANTEN, which suggests that this cloud is part of a giant molecular cloud (GMC).

The molecular emission of Cloud 3  appears mostly located in the region of the optical nebula. The bulk of the molecular gas is concentrated in  an elongated feature extending from  between \radec\ $\sim$ (18$^h$19$^m$05$^s$,\hbox{--11\deg 34\arcmin)} to \radec\ $\sim$ (18$^h$19$^m$,\hbox{--11\deg 39\arcmin)}. At lower declinations, this feature is shifted toward the east, encircling the position of the WR star. The size and morphology of Cloud 3 is in excellent agreement with the optical nebula.   The NANTEN image corresponding to  Cloud 3 shows that the bulk of the molecular gas of this cloud coincides with of the optical nebula and does not extends beyond the region surveyed by the SEST data.

The large extent of Cloud 2 and the morphological coincidence with a region lacking  optical emission towards the north  and west of the nebula speaks in favour of an interpretation in which Cloud 2 is  a  foreground object with respect to WR 113, or is placed at almost the same distance. On the other hand, the good morphological correspondence of Cloud 3 with the inner optical nebula around WR 113 suggests that this structure is physically related to the ionized ring nebula. It ought to be pointed out that along this galactic longitude, considering mean radial velocities of about $\sim$+27.5 \kms\ and $\sim$+35.5  \kms, the galactic velocity field of \citet{bb93}\  (see their fig. 2b) along \hbox{{\it l} $\approx$ 20\gra} predicts near kinematical distances of 1.8 $\pm$ 0.5 kpc  and 2.3 $\pm$ 0.5 kpc for Cloud 2 and Cloud 3, respectively. Uncertainties in these distances were obtained adopting random motions of $\pm$6 \kms\citep{bg78}.  Keepeing this in mind an interpretation  in which Cloud 3 is behind Cloud 2 can not be ruled out. Nevertheless, from the present data we cannot either  confirm or deny a physical relation between Cloud 2 and the optical nebula.   Based on the above, from here onwards we shall concentrate on the analysis of Cloud 3,which is the only cloud clearly  associated with the optical ring nebula around WR 113.

In order to better characterize the kinematical  properties of Cloud 3, in Fig. \ref{fig:serie-1}    a collection of images depicting the \hbox{CO(1-0)} spatial distribution in the velocity  range from +28.9 to +36.9  \kms\ is shown. Every image displays the CO emission averaged over a velocity interval of $\sim$0.9 \kms\ (three individual smoothed  channel maps). The CO emission distribution shown in Fig. \ref{fig:serie-1}   (in contours) is  projected onto the DSSR image of the optical nebula (greyscale). The velocity interval of the individual images is indicated in the lower right corner. In the velocity interval from   +29.8 to +32.5 \kms\ the bulk of the molecular emission is concentrated towards the center of the optical nebula and is projected onto the region of relatively faint emission between the stellar position and the bright optical filaments. This coincidence suggests that the faint optical emission observed towards the center of the nebula arises from ionization of photodissociated molecular gas. The VLA image also shows low radio continuum emission in this area, indicating the presence of ionized gas. From +32.5 \kms\ the molecular gas develops an arc-like morphology, which is better noticed at velocities between $\sim$ +34.3 and +35.2. This morphology remains at 36.0 \kms\ where is still noticed as a hollow patchy anular feature.

\subsection{Ratio $R_{2-1/1-0}$ and excitation temperature}

A powerful tool that help us to determine the physical conditions of molecular  clouds is the $R_{2-1/1-0}$ ratio.         \citet{sak97} classified the molecular gas in three categories: ``very high ratio gas'' (VHRG, corresponding to $1.0 <R_{2-1/1-0}< 1.2$), `` high ratio gas'' (HRG, for $0.7 <R_{2-1/1-0}<1$), and `` low ratio gas'' (LRG, for $R_{2-1/1-0}<0.7$). Values compatible with VHRG denote surfaces of dense clumps irradiated by strong radiation field and PDRs   \citep{gi92}. In this case,  excitation temperatures ($T_{\rm exc}$) greater than  50  K and volume densities  greater than  3$\times$10$^3$ cm$^{-3}$ are expected. Molecular gas with  $R_{2-1/1-0} \simeq$ 1.3 is observed toward \hii\ regions  \citep{cas90,sak94}, which heat its surrounding molecular dense gas. Ratios in the range 0.7 to 1.0 are usually observed in central regions of giant molecular clouds where CO emission originates in collision processes (Castets et al. 1990, Sakamoto et al. 1994). This material is characterized by $T_{\rm exc} \gtrsim$ 20 K and  $n_{\rm H_2}\ \gtrsim$ 1 $\times$ 10$^3$ cm$^{-3}$. Finally, $R_{2-1/1-0}<0.7$  is observed in the outer envelopes of giants molecular clouds. These values are consistent with gas subthermally excited, with $T_{\rm exc} \leq$ 10 K and densities $n_{\rm H_2} <$ 1$\times$10$^3$ cm$^{-3}$. To probe the surface conditions of Cloud 3,   we will analyze the $T_{\rm exc}$-distribution  obtained from the CO(1-0) line, assumed to be optically thick.

\begin{figure}
\centering
\includegraphics[angle=0,width=90mm]{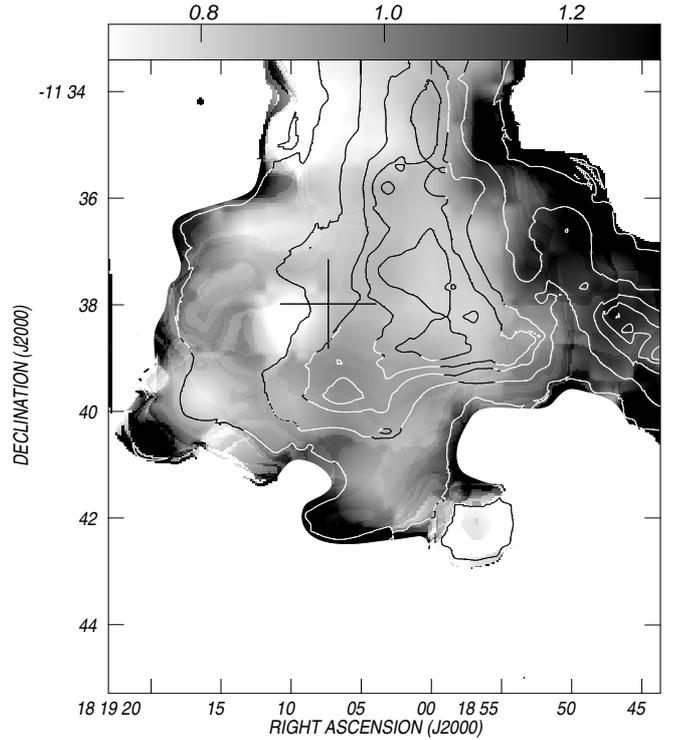}
\caption{Line ratios $R_{2-1/1-0}$ for Cloud 3. The grayscale goes from $\sim$ 0.6 (light gray) to $\sim$ 1.3 (dark gray).   Black contours correspond to the $T_{\rm exc}$ of the $^{12}$CO(1-0) line.  $T_{\rm exc}$ contours are 10,  20, 30, 35, 40, and  50   K.  }
\label{fig:ratio}
\end{figure}

Fig.~\ref{fig:ratio} shows the spatial distribution of the ratio $R_{2-1/1-0}$ obtained for Cloud 3 in grayscale, where 
\begin{equation}
R_{2-1/1-0} = \frac{\int_{\rm v_1}^{\rm v_2}  T_{mb\ (2-1)}\ d{\rm v}}{ \int_{\rm v_1}^{\rm v_2}  T_{mb\ (1-0)}\ d{\rm v}} 
\end{equation}
with v$_1$ = 29 \kms and v$_2$ = 36.5 \kms. The spatial distribution of $\int_{\rm v_1}^{\rm v_2}  T_{mb\ (2-1)}\ d{\rm v}$ and $\int_{\rm v_1}^{\rm v_2}  T_{mb\ (1-0)}\ d{\rm v}$ (order-zero moments)   were obtained using AIPS package.  

 The  $T_{\rm exc}$ distribution of the $^{12}$CO line can be obtained from $T_{peak}\ (^{12}CO) = T^*[J_{\nu}(T_{\rm exc}) - J_{\nu}(T_{bg})]$\ \citep{d78}, where $T_{peak}$ is the peak temperature of the $^{12}$CO(1-0) line, $T^*=h\nu /k$, with $\nu$ the rest frequency of the $^{12}$CO(1-0) line, and  $J_{\nu}(T)={(\exp(T^*/T) - 1)}^{-1}$. We adopt a background temperature $T_{bg}$ $\sim$ 2.7 K. Using this equation, assuming Gaussian profiles for the CO(1-0) line, and combining the order-zero moment map (i.e., integrated areas) with the order-two moment maps (i.e., velocity dispersion), we obtain the $T_{\rm exc}$ distribution map, which is depicted in Fig. \ref{fig:ratio} with black contours.

An inspection of Fig.~\ref{fig:ratio} shows values of  $R_{2-1/1-0}$  between 0.5 and 1.2 along Cloud 3.  It is worth to mention that values of $R_{2-1/1-0}$ higher than 1.2 in the border of the CO cloud can not be taken into account due to their large uncertainties.     Two small regions at   18$^h$19$^m$07$^s$ $<$ RA $<$ 18$^h$19$^m$13$^s$ and  --11\deg 37.\arcmin5  $<$  DEC $<$  --11\deg 39\arcmin, and at  18$^h$19$^m$05$^s$ $<$ RA $<$ 18$^h$19$^m$12$^s$ and   --11\deg 33.\arcmin5  $<$  DEC $<$  --11\deg 35.\arcmin5  exhibit  values of $R_{2-1/1-0}$  between $\sim$ 0.4  to $\sim$ 0.7, probably arising in  spots of quiescent dark molecular gas. The rest of the molecular cloud shows higher values (between $\sim$ 0.8 and $\sim$ 1.1), which   suggest that the molecular gas of Cloud 3  is being irradiated by an external UV field and  might be indicative of  the presence of PDRs along Cloud 3. A similar conclusion can be dropped from the $T_{\rm exc}$-analysis.  Excitation temperatures of about $\sim$ 50 K are observed towards the brightest part of Cloud 3, which are  higher than expected inside molecular cores if only cosmic ray ionization is considered as the main heating source \hbox{($T$ $\sim$ 8 - 10 K,   \citealt{vdt00})}. Excitation temperatures of about $\sim$ 30 K are observed toward molecular clouds located at the edges of evolved \hii\ regions ({\it bright rimmed clouds}), which implies that additional heating processes, like photoionization, are present close to these clouds \citep{U09}. We conclude that Cloud 3 is  being externally heated through the photoionisation of its surface layers as a consequence of its proximity to neighbouring ionizing stars.

\subsection{Physical parameters of the molecular cloud}

Table~\ref{table:prop} lists the main physical parameters of Cloud 3, obtained from the $^{12}$CO(1-0) observations. The table includes angular sizes, velocity ranges, systemic velocities, mean H$_2$ column densities, and molecular masses. Mean H$_2$ column densities were derived using the empirical relation between the integrated emission $I_{\rm CO}$ (= $\int T_{mb}\ d{\rm v})$ and N$_{\rm H_2}$. We adopted  $N({\rm H_2}) = (1.9\ \pm\ 0.3)\ \times\ 10^{20}\ I_{\rm CO}$ \ (cm$^{-2}$)  \citep{ d96,sm96}.  The total molecular mass  was calculated using  $M_{\rm H_2} =  (m_{sun})^{-1}\  \mu\ m_H\ \sum\ \Omega\ N({\rm H_2})\ d^2$  (M$_{\odot}$), where  $m_{sun}$ is the solar mass ($\sim$ 2 $\times$ 10$^{33}$ g),    $\mu$ is the mean molecular weight, assumed to be equal to 2.8 after allowance of a relative helium abundance of 25\% by mass \citep{Y99},  $m_{H}$ is the hydrogen atom mass   ($\sim$ 1.67 $\times$ 10$^{-24}$ g), $\Omega$ is the solid angle subtended by the CO feature  in ster, and $d$ is the  distance. The mean volume density ($n_{\rm H_2}$) of Cloud 3  can be derived from the ratio of its molecular mass and its volume considering an approximately spherical geometry with a radious of about 2.3 pc.


\begin{table}
\caption{Main physical parameters of Cloud 3.}
\label{table:prop}
\begin{center}
\begin{tabular}{ l  c}
\hline\hline
Parameter  &  Value\\
\hline
Angular size  (\arcmin)          & $\sim$ 7   \\
{\bf Linear size  (pc)  }        &  {\bf 4.0 $\pm$ 1.2}    \\
Velocity range (\kms)          & $\sim$ +30 , +36 \\
Systemic velocity (\kms)            & $\sim$ 35.5 $\pm$ 1.0   \\
H$_2$ column density  $N_{\rm H_2}$ (10$^{21}$ cm$^{-2}$)   &   4.1 $\pm$ 0.6\\
Molecular mass $M_{\rm H_2}$  (M$_{\odot}$)       &  $\sim$1600 $\pm$ 500    \\
$n_{\rm H_2}$ (molecules cm$^{-3}$)                &  $\sim$ 450     \\


\hline

\end{tabular}
\end{center}
\end{table}

\section{IR counterparts  and photodissociated regions}\label{ir}

\begin{figure}[h!]
\centering
\includegraphics[angle=0,width=93mm]{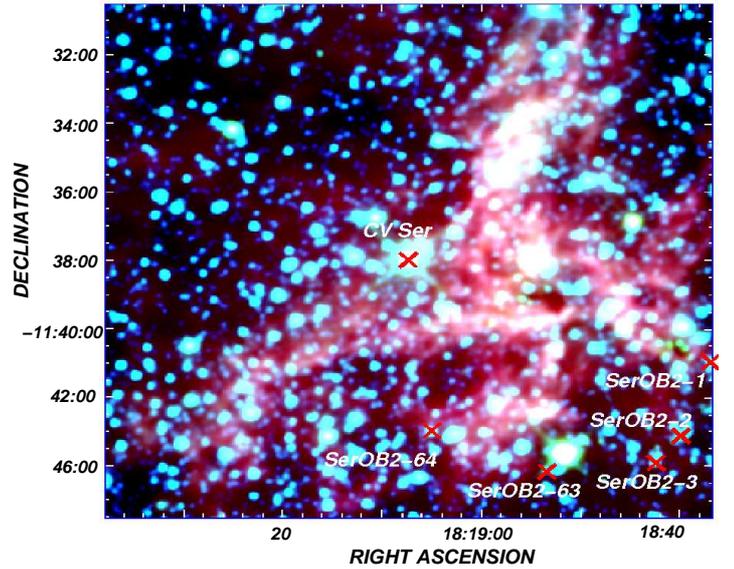}
\caption{Superposition of the MIR emissions at 3.4 $\mu$m (blue), 4.6 $\mu$m (green), and 12.0 $\mu$m (red). The positions of WR 113 and the Ser\,OB2 members  are indicated by red crosses.}
\label{wise}
\end{figure}

 In Fig.~\ref{wise} we show the  mid infrarred (MIR)  emission at 3.4 $\mu$m (blue), 4.6 $\mu$m (green), and 12.0 $\mu$m (red) in the region  of the inner ring nebula. The 3.4 $\mu$m and 12 $\mu$m filters include prominent PAH emission features and the 4.6 $\mu$m filter measures the continuum emission from very small grains \citep{wr10}. The  emision at 12.0 $\mu$m resembles  the observed emission at 8.0 $\mu$m (see Fig.~\ref{opt1}). The emission at  4.6 $\mu$m is barely detected towards some stellar sources while the emission at 3.4 and 12  \micr\ delineates clearly the inner optical ring.

\begin{figure*}
\centering
\includegraphics[angle=0,width=160mm]{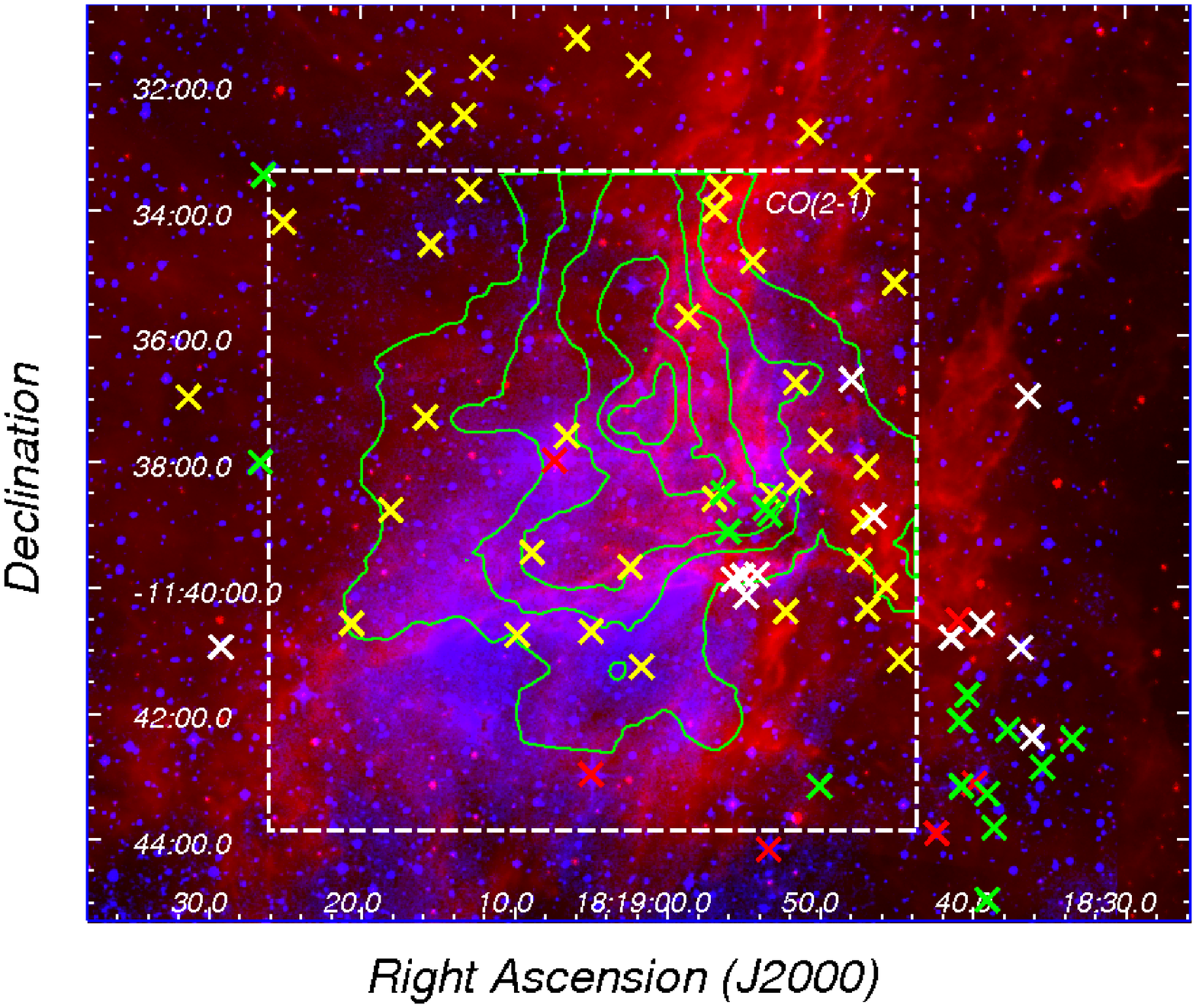}
\caption{Overlay of the 8$\mu$m-image (in red) , the optical (DSS2\,R,
 in blue), and millimetre (CO(2-1), green contours) images of the
region. The yellow crosses indicate the position of the 2\,MASS pointo
sources with IR excess, while the white and green crosses the IRAC
class I and class II sourecs, respectively. Finally, the red crosses
indicate the position of CV\,Ser and Ser\,OB-1,-2,-3,-63 and -64.  }
\label{spitzer-yso}
\end{figure*}

 Fig.~\ref{spitzer-yso} shows the IR emission distribution from the Spitzer-IRAC 8.0 $\mu$m image (red), the optical DSS\,2R image (blue), and the CO(2-1) contours over a field larger than the region surveyed with SEST. The emission in 8.0 $\mu$m, mainly originated in the  strong 7.7 and 8.3 $\mu$m PAH features, shows a concentration of material along two elongated almost perpendicular regions  placed at $R.A. \sim$ 18$^h$18$^m$55$^s$, extending from $Decl.$ $\sim$ --11\deg 40\arcmin\ to --11\deg 30\arcmin, and at  $Decl.$ $\sim$ --11\deg 40\arcmin, from $R.A.$ $\sim$ 18$^h$19$^m$30$^s$ to  18$^h$18$^m$50$^s$. As we described in Section~\ref{sc}, the optical inner bright filament appears bordering the MIR emission.  Fig.~\ref{spitzer-yso}   also shows the  diffuse optical emission north and south of the inner optical filament. Both emissions, optical and MIR, enclose the molecular cloud towards the south and west.\

   The presence of PAH emission, which is typical of photodissociation regions (PDRs), at the interfase between the ionized and molecular gas suggests that the molecular gas is being photodissociated by the UV photons emitted by the massive stars \citep{va10}.  For this reason and according to the morphology of the PAH features observed in Figs.~\ref{wise} and~\ref{spitzer-yso}, and the optical and molecular emissions (see Fig.~\ref{fig:serie-1}), we can confirm the presence of a PDR located at $Decl.$ $\sim$ --11\deg 41\arcmin, from $R.A.$ $\sim$ 18$^h$18$^m$50$^s$ to  18$^h$19$^m$15$^s$ which could be generated mainly by the action of the photo-dissociating far-ultraviolet (6 eV$<h\nu<$ 13.6 eV) photons of Ser\,OB2-1,-2,-3,-63, and -64.

\section{Star formation}\label{st}

To investigate the presence of protostellar candidates in the region  we used data from the available  MSX, 2\,MASS, IRAS, and {\it Spitzer}-IRAC  point source catalogs, in a region of about 10\arcmin\ in size centered on the position of the WR star.

\begin{figure}[h!]
\centering
\includegraphics[width=93mm]{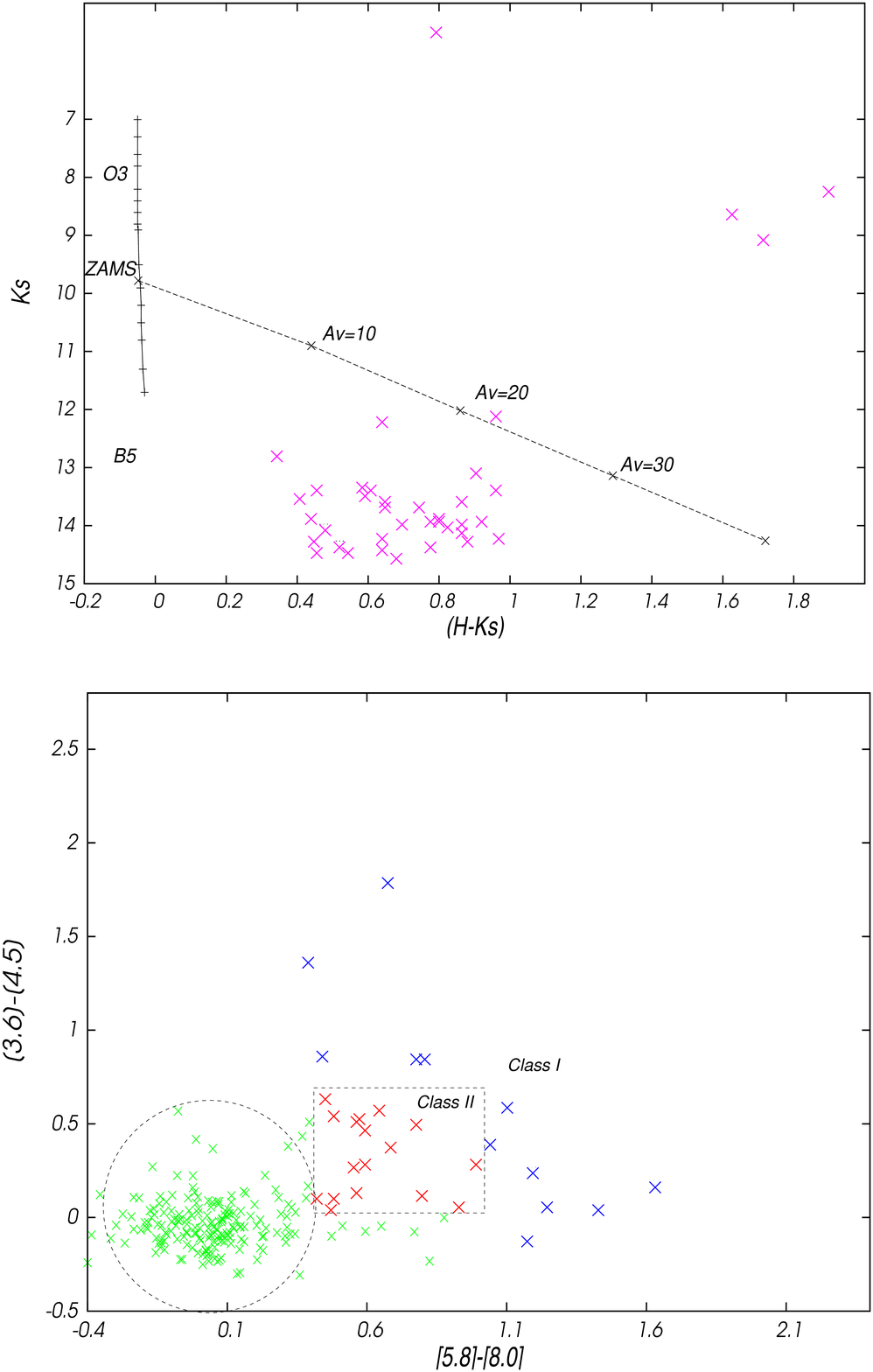}
\caption{{\it Upper panel}: Color-magnitude diagram for the 2MASS sources with IR excess. {\it Lower panel}: Color-color diagram for the {\it Spitzer} sources around the region of study.  }
\label{ysos}
\end{figure}

    From an original set of $\sim$18000 2\,MASS sources, we selected those with high photometrical quality in the $J$, $H$, and $Ks$ bands. Following \citet{co05} we look for candidate YSOs using the parameter \hbox{$q$ = ($J$ - $H$) - 1.83 $\times$ ($H$ - $K_s$)}.  This formula allows one to  discriminate between giant and main sequence stars, and sources with IR excess. Sources with $q <$ --0.15 are classified as objects with infrared excess, revealing the presence of dusty circumstellar envelopes, i.e. candidate YSOs. After applying this criterion, we found 39 sources with IR excess. Fig.~\ref{ysos} (upper panel) displays the color-magnitude (CM) diagram of these sources. The ZAMS from O3 to B5 type stars is displayed at the left of the diagram, assuming a distance of 2.0 kpc. Almost all the sources with IR excess  present visual absorption larger than 10 mag. Table~\ref{tabla3} summarizes the main parameters of these sources, i.e. galactic coordinates, designation, and $J$, $H$, and $Ks$ magnitudes.\

\onltab{3}{
\begin{table*}
\caption{YSO candidates projected over the region.} 
\label{tabla3}
\centering
\begin{tabular}{lllcccc}
\hline\hline
 \multicolumn{7}{c}{2\,MASS} \\
   Name          & \multicolumn{2}{c}{\radecb}    & $J$ [mag]   & $H$ [mag]   & $K_s$ [mag] & \\
\hline  
18190736-1137591 & 18:19:7 & -11:37:59.1312& 7.016 & 6.283 & 5.492 &  \\
18190884-1139265 & 18:19:8 & -11:39:26.5932& 15.859 & 14.684 & 13.888 & \\
18185686-1134000 & 18:18:56 & -11:34:0.0192& 13.448 & 10.152 & 8.255 & \\
18185863-1135413 & 18:18:58 & -11:35:41.3448& 14.492 & 13.956 & 13.547 & \\
18190584-1131161 & 18:19:5 & -11:31:16.1292& 16.194 & 15.234 & 14.556 & \\
18185132-1138198 & 18:18:51 & -11:38:19.8348& 14.913 & 13.997 & 13.395 & \\
18184741-1139346 & 18:18:47 & -11:39:34.668& 14.347 & 11.825 & 9.379 & \\
18185437-1134484 & 18:18:54 & -11:34:48.4104& 15.831 & 14.457 & 13.591 & \\
18185653-1133388 & 18:18:56 & -11:33:38.808& 16.310 & 14.856 & 13.937 & \\
18191623-1131588 & 18:19:16 & -11:31:58.8& 15.439 & 14.014 & 13.106 & \\
18191547-1132478 & 18:19:15 & -11:32:47.832& 15.336 & 14.708 & 14.256 & \\
18191209-1131439 & 18:19:12 & -11:31:43.9248& 13.528 & 13.162 & 12.820 & \\
18184689-1140203 & 18:18:46 & -11:40:20.37& 16.275 & 15.167 & 14.198 & \\
18191326-1132294 & 18:19:13 & -11:32:29.4576& 16.161 & 14.827 & 13.961 & \\
18191548-1134327 & 18:19:15 & -11:34:32.7216& 15.955 & 14.709 & 13.938 & \\
18184570-1139594 & 18:18:45 & -11:39:59.4864& 13.532 & 10.775 & 9.065 & \\
18190241-1139407 & 18:19:2 & -11:39:40.7556& 13.711 & 12.860 & 12.218 & \\
18190655-1137351 & 18:19:6 & -11:37:35.148& 14.757 & 13.902 & 13.322 & \\
18192507-1134108 & 18:19:25 & -11:34:10.8876& 12.806 & 10.280 & 8.657 & \\
18184479-1141092 & 18:18:44 & -11:41:9.2148& 15.483 & 14.647 & 13.955 & \\
18185316-1138319 & 18:18:53 & -11:38:31.9632& 15.598 & 14.851 & 14.331 & \\
18184510-1135094 & 18:18:45 & -11:35:9.4632& 14.612 & 13.099 & 12.139 & \\
18191292-1133406 & 18:19:12 & -11:33:40.6296& 14.524 & 13.864 & 13.407 & \\
18185691-1138354 & 18:18:56 & -11:38:35.4084& 14.700 & 14.329 & 13.891 & \\
18184716-1138575 & 18:18:47 & -11:38:57.5988& 15.651 & 14.994 & 14.453 & \\
18193130-1136578 & 18:19:31 & -11:36:57.8808& 15.991 & 15.031 & 14.388 & \\
18190185-1131418 & 18:19:1 & -11:31:41.8692& 16.039 & 15.142 & 14.363 & \\
18190169-1141155 & 18:19:1 & -11:41:15.5688& 15.338 & 14.318 & 13.671 & \\
18184726-1133344 & 18:18:47 & -11:33:34.4772& 15.087 & 14.519 & 14.044 & \\
18191809-1138460 & 18:19:18 & -11:38:46.0824& 15.443 & 14.408 & 13.668 & \\
18185062-1132452 & 18:18:50 & -11:32:45.294& 16.203 & 15.118 & 14.242 & \\
18184694-1138052 & 18:18:46 & -11:38:5.2584& 15.518 & 14.927 & 14.469 & \\
18185224-1140231 & 18:18:52 & -11:40:23.1708& 14.921 & 14.073 & 13.484 & \\
18190983-1140452 & 18:19:9 & -11:40:45.228& 16.219 & 14.999 & 14.138 & \\
18184996-1137399 & 18:18:49 & -11:37:39.9936& 15.938 & 14.338 & 13.379 & \\
18191577-1137174 & 18:19:15 & -11:37:17.4504& 15.883 & 14.867 & 14.223 & \\
18185159-1136448 & 18:18:51 & -11:36:44.8344& 15.644 & 14.755 & 13.955 & \\
18190499-1140419 & 18:19:4 & -11:40:41.9808& 15.997 & 14.832 & 14.012 & \\
18192064-1140341 & 18:19:20 & -11:40:34.1004& 15.215 & 14.217 & 13.566 & \\
\hline  
\hline
 \multicolumn{7}{c}{IRAC (Class I)} \\
 Name        & \multicolumn{2}{c}{\radecb}    & $[3.6]$ [mJy]  & $[4.5]$ [mJy]  & $[5.8]$ [mJy] & $[8.0]$[mJy]    \\
\hline
G018.8888+01.8312 & 18:18:47 & -11:36:41.6088 & 1.389000 & 0.923700 & 2.600000 & 5.412000\\
G018.8152+01.8312 & 18:18:39 & -11:40:34.9572 & 4.999000 & 6.915000 & 9.280000 & 10.660000\\
G018.8190+01.8261 & 18:18:40 & -11:40:31.3932 & 2.304000 & 5.135000 & 10.280000 & 8.187000\\
G018.8158+01.8223 & 18:18:41 & -11:40:48.1116 & 8.123000 & 11.270000 & 15.860000 & 18.620001\\
G018.8550+01.7845 & 18:18:54 & -11:39:48.0924 & 19.340000 & 15.520000 & 26.340000 & 43.950001\\
G018.8514+01.7792 & 18:18:54 & -11:40:8.3856 & 0.895400 & 1.261000 & 2.747000 & 2.281000\\
G018.8540+01.8193 & 18:18:46 & -11:38:52.0728 & 4.180000 & 3.822000 & 4.604000 & 6.690000\\
G018.8626+01.8707 & 18:18:36 & -11:36:57.4668 & 2.881000 & 9.610000 & 20.260000 & 21.110001\\
G018.8565+01.7810 & 18:18:55 & -11:39:49.2012 & 2.993000 & 2.224000 & 6.030000 & 15.150000\\
G018.8569+01.7784 & 18:18:55 & -11:39:52.3584 & 3.836000 & 2.565000 & 4.646000 & 8.199000\\
G018.7823+01.8288 & 18:18:36 & -11:42:23.4252 & 5.035000 & 5.525000 & 6.159000 & 9.447000\\
G018.9059+01.6493 & 18:19:29 & -11:40:56.3916 & 1.625000 & 0.925700 & 1.329000 & 2.185000\\
\hline  
 \multicolumn{7}{c}{IRAC (Class II)} \\
\hline
G018.7887+01.7997 & 18:18:43 & -11:42:52.6608 & 11.310000 & 10.780000 & 10.240000 & 8.016000\\
G018.7500+01.7978 & 18:18:39 & -11:44:58.5132 & 1.633000 & 1.678000 & 1.464000 & 1.380000\\
G018.8687+01.7831 & 18:18:55 & -11:39:6.9192 & 6.715000 & 7.711000 & 9.600000 & 8.156000\\
G018.7872+01.8240 & 18:18:37 & -11:42:16.1028 & 3.466000 & 3.565000 & 3.959000 & 3.744000\\
G018.7660+01.8084 & 18:18:38 & -11:43:49.5444 & 12.800000 & 12.590000 & 11.850000 & 11.430000\\
G018.7741+01.8274 & 18:18:35 & -11:42:51.7788 & 3.781000 & 3.821000 & 3.752000 & 4.299000\\
G018.7748+01.8111 & 18:18:39 & -11:43:17.2596 & 1.671000 & 1.770000 & 1.929000 & 1.675000\\
G018.7760+01.7794 & 18:18:46 & -11:44:7.3068 & 2.620000 & 2.677000 & 2.298000 & 1.841000\\
G018.7769+01.8381 & 18:18:33 & -11:42:24.696 & 2.354000 & 2.556000 & 2.122000 & 2.144000\\
G018.9792+01.7830 & 18:19:80 & -11:33:16.7724 & 10.160000 & 9.544000 & 5.075000 & 2.799000\\
G018.7823+01.8288 & 18:18:36 & -11:42:23.4252 & 5.035000 & 5.525000 & 6.159000 & 9.447000\\
G018.7723+01.7188 & 18:18:58 & -11:46:2.0496 & 2.182000 & 2.357000 & 1.775000 & 0.922700\\
\hline
\end{tabular}
\end{table*}
}


 For the case of IRAS and MSX sources, we search for candidate YSOs following  the criteria of    \citet{ju92} and \citet{lu02}, respectively. In the case of  the IRAS catalog, we do not find any sources satisfying the mentioned criteria into the selected region, while in the case of the MSX catalog, we find only one source at \radec\ = \hbox{(18$^h$18$^m$56.$^s$4, --11\gra 34\arcmin0.\arcsec5)}, candidate to compact \hii\ region.\

  In order to complete the search for candidate YSOs around the optical nebula, we use the GLIMPSE 3D (2007-2009) catalog to perform a photometric study in the region. Considering only sources detected in the four {\it Spitzer}-IRAC bands, we found 8477 sources. To investigate the evolutionary stage of these sources, we have analyzed their location in a color-color diagram  (Fig. \ref{ysos}, lower panel), following the color criteria from \citet{all04}. As expected, most of sources seem to cluster around (0,0). This region of the diagram contain mostly  background/foreground stars  and Class III sources with no intrinsic IR excess. Sources in  red  occupy the Class II region (\hbox{0 $\lesssim$ [3.6]-[4.5] $\lesssim$0.7}, \hbox{0.4 $\lesssim$[5.8]-[8.0] $\lesssim$1}), and their IR excess might be produced by accretion disks around the stellar object. Sources in the domain of Class I objects (\hbox{0.7 $\lesssim$ [3.6]-[4.5]}, \hbox{1 $\lesssim$[5.8]-[8.0]})   are displayed in blue. In this case, the IR excess  originates in circumstellar envelopes  around the young stellar object.

 In Fig.\ref{spitzer-yso} we display the position of the candidate YSOs   mentioned above.  Many of them  are projected outside Cloud 3, which suggests that these sources are background/foreground objects  unrelated to the region of study. Among the candidate YSOs projected onto Cloud 3, 22 are 2MASS sources, while only six and four are Class I and Class II, respectively. Unlike the 2MASS sources, which are spread  onto Cloud 3, Class I and Class II sources appear concentrated in two spots near   \radec\ = \hbox{(18$^h$18$^m$52$^s$,--11\gra 38\arcmin45\arcsec)}. This location  is almost coincident with a region of high  optical absorption and a clump of molecular gas better noticed at velocities between +33.4 and +35.2 \kms (see Fig. \ref{fig:serie-1}). This coincidence shows clearly that star formation is active in this region. The small value of the molecular mass ($\sim$1600 \msun ) suggests that this cloud is the remnant of a larger parental cloud that has been photoionized and dissociated by the emerging young stars. To investigate whether these  candidate Class I and Class II sources   may have been triggered by the expansion of the molecular gas in a ``collect and collapse'' scenario we applied the analytical model of \citet{wi94}.  According to this model expanding nebulae compress gas between the ionization and the shock fronts, leading to the formation of  molecular cores where new stars can be embedded. Using the Whitworth et al.'s formulae   for the case of expanding \hii\ regions, we derived the time when the  fragmentation may have occurred, $t_{frag}$, and  the size of the \hii\ region at $t_{frag}$, $R_{frag}$, which are given by $t_{\rm frag} [10^6 yr]\ =\ 1.56 a_2^{4/11}\ n_3^{-6/11}\ N_{49}^{-1/11}$ and $R_{\rm frag} [pc]\ =\ 5.8 a_2^{4/11}\ n_3^{-6/11}\ N_{49}^{1/11}$, where $a_2$ is the sound velocity in units of 0.2 \kms,  $n_3 \equiv  n_{H_2}/1000$, and $N_{49}\equiv N_{Lyc}^*/10^{49}$. Adopting 0.3 \kms for the sound velocity (corresponding to temperatures of 30–50 K in the surrounding molecular clouds), we obtained $t_{\rm frag}$ $\sim$ 1.2 $\times$ 10$^6$ yr, and R$_{\rm frag}$ $\sim$ 15 pc. Considering that $t_{\rm frag}$ and $R_{\rm frag}$ are higher than the dynamical age of Cloud 3 ($t_{\rm dyn}$ $\sim$ 1.0 $\times$ 10$^5$ yr; see Sect 6) and its radious ($R_{\rm cloud\ 3}$ $\sim$ 2.3 pc) we  conclude that the fragmentation at the edge of Cloud 3 is doubtful.
Nevertheless, the clustered aspect of these sources make them  excellent candidates for investigating star formation with high angular resolution observations.

\begin{figure*}
\centering
\includegraphics[angle=0,width=155mm]{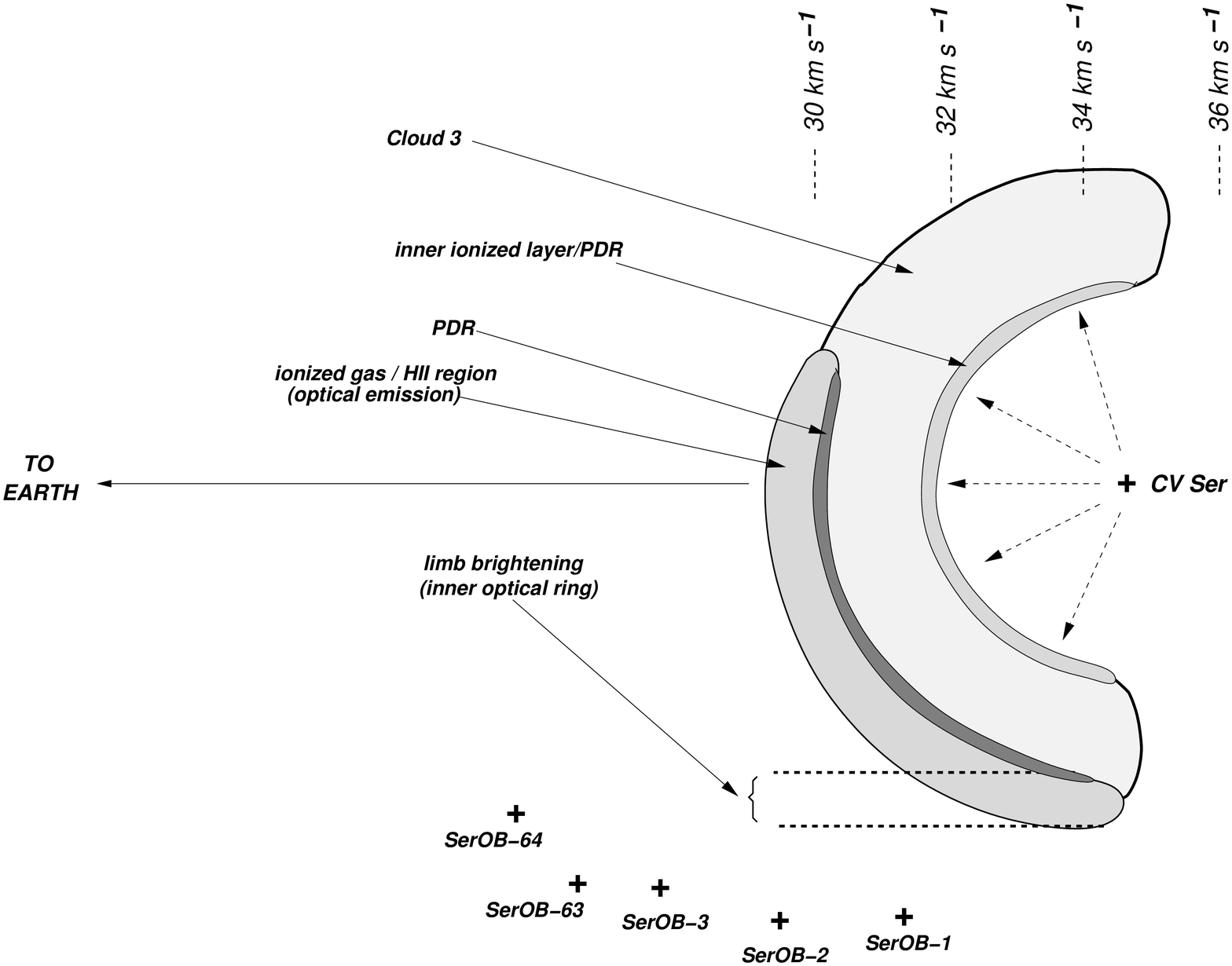}
\caption{Sketch of the model used to explain the characteristics of the molecular and ionized gas.   }
\label{fig:dibujo}
\end{figure*}

\section{Scenario}\label{scenario}

 As mentioned in Section~\ref{sc}, \citet{gr84} and \citet{er95} (using optical lines), and \citet{c02} (using radio continuum data), explained  the inner optical bright filament and its radio counterparts as the result of action of the UV field and the strong stellar winds of the star WR\,113.\ An alternative scenario can be proposed taking into account the kinematical properties of Cloud 3, and the location of the WR star and SerOB2-1, -2, -3, -63, and -64. 

Along the velocity interval from  +32.5 to +36 \kms, WR 113 appears projected onto a region of faint molecular\   emission, which suggests that the star is interacting with the molecular cloud. The relatively large velocity range of Cloud 3 ($\sim$ 7 \kms) suggests that expanding motions are  present in the molecular gas.   The classical wind-blown expanding shell scenario predicts that the surrounding gas expands spherically around the star. Then, if the molecular emission is a  two-dimensional projection of an expanding spherical shell, it should  appear in the data cube   first as a blueshifted pole (approaching cap), then as a growing-decreasing circular  ring, and finally as a redshifted pole (receding cap), with the powering star placed at the center or close to it. According to Fig. \ref{fig:serie-1},  the approaching section of this shell is present, although  the receding part is missed. Note also  that the brightest  optical ring coincides with the outer border of Cloud 3, indicating that this cloud is also being  photodissociated by an external  source, possibly located towards the southwest. Bearing in mind the location of  SerOB-1, -2, -3, -63 and -64 (see Table \ref{tabla1}) and their linear distance to Cloud 3 (1 - 2  pc, taking into account a distance of 2.0 kpc), we speculate that these stars are responsible for the inner optical ring.    The diffuse optical emission observed north of the bright optical ring, in the region between 18$^h$18$^m$50$^s$ $<$ RA $<$ 18$^h$19$^m$20$^s$ and   --11\deg 36\arcmin  $<$  DEC $<$  --11\deg 40\arcmin\         might arise from the photodissociation of the molecular gas at the surface of Cloud 3 that faces the Earth, probably at lower radial velocities (between +30 and +32 \kms), since  we indeed detect optical  emission. The diffuse optical emission at the south of the optical rim, between  --11\deg 40\arcmin  $<$  DEC $<$  --11\deg 42\arcmin    probably originates in the dissociation of the small molecular clump at  \radec\ = \hbox{(18$^h$19$^m$3$^s$,--11\gra 41\arcmin30\arcsec)}, with  velocities  between +34.3 and +36.0 \kms.

In Fig. \ref{fig:dibujo} we present a  side view of a simple sketch of the scenario proposed to explain the characteristics of the molecular and ionized gas. Cloud 3 is a half-shell which expands around CV Ser as the result of the wind mechanical energy injected into the ambient medium  by the binary system. The stars SerOB2-1, -2, -3, -63, and -64, belonging to SerOB2, photodissociate and ionize the southern  and central region of the outer face of Cloud 3 giving rise to the diffuse  emission observed at optical wavelengths.  According to this scenario,  the intensity of the optical emission is expected to rise as the thickness  of the emitting ionized region in the line of sight increases. Then, the maximum of the optical emission should be observed toward the edge of the molecular cloud ({\it limb brightening}), which explain the location of the inner optical ring at the southern and western borders of Cloud 3 (see Fig. \ref{fig:dibujo}). The PAH emission is originated in the PDR at the interface between the ionized and molecular gas.  The UV field of CV Ser ionizes the inner face of Cloud 3, although  the optical emission is probably mostly absorbed by the dust linked to the molecular gas in the line of sight.\

   We estimated an  expansion velocity, $V_{\rm exp}$ = 6 \kms. Similar expansion velocity values were derived for the expanding shells linked to WR stars, e.g. $\sim$7.4 \kms (\citet{va09}, $\sim$9 \kms (\citet{ca09}), 8$\pm$1 \kms (\citet{va10}).  Adopting an expansion velocity of 6 \kms, the dynamical age of the half-shell, according to wind blown bubble models  can be calculated using $t_{\rm dyn}\ =\ 0.5 \times 10^6\ \times\  \frac{R}{V_{\rm exp}}$ \citep{mc83,lc99}, where $R$ is the radius of the bubble (pc). We obtain   $t_{\rm dyn}$ $\approx$ 2 $\times$ 10$^5$ yr, which is in agreement, within the errors, with the dynamical age obtained by \citet{c02} using radio continuum data.

From the value of the molecular mass and the expansion velocity, we determined a kinetic energy  $E_{\rm kin}$ = 6$\times$10$^{47}$ erg and a momentum $P$ = 9.6$\times$10$^3$ \msun\ \kms\ for the molecular cloud. To calculate the mechanical energy injected by CV\,Ser, we shall assume typical values for the stellar wind of a WC\,8 star ($\dot{M}$ = 2.5$\times$10$^{-5}$ \msunyr,\ $v_\infty$ = 1890 \kms; \citealt{nl00}), and for its O9V type star companion ($\dot{M}$ = 4.0$\times$10$^{-6}$ \msunyr\ and $v_{\infty}$ = 1800 \kms; \citealt{snc02}). We obtain for CV\,Ser a  mechanical luminosity of  $L_{\rm w}\simeq$3.2 $\times$ 10$^{37}$ erg s$^{-1}$. Considering a lifetime for the binary system equal to the age of Ser\,OB2  (5 $\times$ 10$^6$ yr; \citealt{f00}), the mechanical energy injected by the binary system into the interstellar medium is $E_{\rm w} \simeq$ 5.1$\times$10$^{51}$ erg. Thus, the energy conversion efficiency  is $\frac{E_{\rm kin}}{E_{\rm w}} \sim0.01\%$. In other words, only a small fraction of  the mechanical energy released by  CV\,Ser is needed to account for  the kinetic energy of the expanding molecular gas. It is useful to stress that conversion efficiencies  of the order of 2-5 $\%$ were reported in  interstellar bubbles \citep{ca01,cp01,cacps03,ca09}. Further, theoretical models and numerical simulations also suggest that in wind bubbles  only a few percent of the injected mechanical energy will be converted into kinetic energy of the expanding gas   \citep{W77,k92,ar07}.

The derived values of energy conversion efficiencies are similar to estimates for other stellar wind bubbles (e.g., \citealt{ca06}). They are in agreement with predictions from recent numerical simulations from \citealt{fre03} and \citep{fre06}. The analysis by freyer et al. takes into account the action of the stellar wind and the ionizing flux from stars of 35 and 60 \msun and find that $\frac{E_{\rm kin}}{E_{\rm w}}$ is in the range 0.10-0.04.  In addition, according to our model Cloud 3 subtends a small solid angle ($\lesssim$ 2$\pi$) which  reduces the wind  mechanical energy available to it. Based on the above derived figures and considerations, the stellar winds of  CV\,Ser are very capable of  providing the kinetic energy of Cloud 3.  Similarly, the momentum injected by the strong stellar wind of CV\,Ser to the ISM, taking into account $\dot{M}$, $v_\infty$ for the binary system is  $P_{\rm CV Ser}$ $\approx$ 2.7 $\times$ 10$^5$ \hbox{\msun\  \kms}, which yields to a conversion  efficiency  $\frac{P}{P_{\rm CV Ser}}\sim$ 4$\%$. The estimated value corresponds neither to the energy nor to the momentum conserving models. 
We believe that the existence of the aproaching molecular shell strongly affects
the energy and momentum efficiencies.\

 \section{Summary} 
 
   With the aim  of investigating  the molecular gas around CV\,Ser  and better understand the interstellar scenario in the environs of the binary system, we analysed SEST  $^{12}$CO($J=1-0$) (HPBW = 44$''$)  and ($J=2-1$) (HPBW = 22$''$) data, and  complementary NANTEN $^{12}$CO(1-0) data. The MIR data at 4.5 to 12 $\mu$m, along with {\it Spitzer} images allowed us to make a more comprehensive study of the dust around the nebula.

  The more important aspects of this study can be summarized as follows:

\begin{itemize}

\item A molecular cloud  in the velocity interval from $\sim$ +30 to +37 \kms (Cloud 3) was found to be associated with  the optical ring nebula around the binary system CVSer. Morphological and kinematical properties indicate that this  cloud is a half shell  expanding around CVSer.\\

\item The strong stellar winds of CV\,Ser are the main responsibles for shaping and for the kinematics of the molecular shell.  \\

\item The inner optical ring is placed along the outer edge of the molecular cloud. This suggests that the stars Ser\,OB2-1,-2,-3, -63 and -64, belonging to SerOB2 and located 1 - 2 pc far from Cloud 3, ionize the outer face of the molecular cloud originating the optical ring nebula  and  the diffuse optical emission observed towards the center of the cloud. The observed PDR located between the molecular cloud and the ionized gas is also   generated by the action of these stars.\\

\item A collection of candidate YSOs are detected towards the molecular cloud. Two small clusters of Class I and Class II objects are projected onto a region of high optical absorption, coincident with a small clump of molecular gas, indicating that star formation is active in this region. Although analytical models indicate that a triggered star formation scenario is {\bf doubtful} more studies are necessary to  shed some light on this  issue .

\end{itemize} 

\begin{acknowledgements}
We acknowledge the anonymous referee of her/his comments. J.V. acknowledges the hospitality of the Astronomy Department of Universidad de Chile during his stay in Santiago, Chile. We are grateful to Dr. N. Mizuno for providing us the NANTEN data. This project was partially financed by CONICET of Argentina under project 
PIP 2488/09,  UNLP under project 11/G093, and ANPCyT under project PICT 
903/08. M.R. is supported by the Chilean {\sl Center for Astrophysics}
FONDAP No. 15010003. M.R. wishes to
acknowledge support from FONDECYT (Chile) grant No. 1080335.

This research has made use of the NASA/ IPAC Infrared Science Archive,
which is operated by the Jet Propulsion Laboratory, California Institute
of Technology, under contract with the National Aeronautics and Space
Administration.
 This publication makes use of data products from the Two Micron All Sky
Survey, which is a joint project of the University of Massachusetts and
the Infrared Processing and Analysis Center/California Institute of
Technology, funded by the National Aeronautics and Space Administration
and the National Science Foundation.
\end{acknowledgements}

\bibliographystyle{aa}
\bibliography{bibliografia-wr113}
 
\IfFileExists{\jobname.bbl}{}
{\typeout{}
\typeout{****************************************************}
\typeout{****************************************************}
\typeout{** Please run "bibtex \jobname" to optain}
\typeout{** the bibliography and then re-run LaTeX}
\typeout{** twice to fix the references!}
\typeout{****************************************************}
\typeout{****************************************************}
\typeout{}

}

\end{document}